\documentclass[osajnl,twocolumn,showpacs]{revtex4}
\usepackage{here}
\usepackage{graphicx}
\graphicspath{{pict/}{}}

\usepackage{bm}

\newcommand{\be}{\begin{equation}}
\newcommand{\ee}{\end{equation}}

\begin{document}

\title{Discrete surface solitons in semi-infinite binary waveguide arrays}

\author{Mario I. Molina}

\affiliation{Departamento de F\'{\i}sica, Facultad de Ciencias,
Universidad de Chile, Santiago, Chile}

\author{Ivan L. Garanovich}
\author{Andrey A. Sukhorukov}
\author{Yuri S. Kivshar}

\affiliation{Nonlinear Physics Centre and Centre for
Ultrahigh-bandwidth Devices for Optical Systems (CUDOS), Research
School of Physical Sciences and Engineering, Australian National
University, Canberra, ACT 0200, Australia}


\begin{abstract}
We analyze discrete surface modes in semi-infinite binary waveguide
arrays, which can support simultaneously two types of discrete
solitons. We demonstrate that the analysis of linear surface states
in such arrays provides important information about the existence of
nonlinear surface modes and their properties. We find numerically
the families of both discrete surface solitons and nonlinear Tamm
(gap) states and study their stability properties.
\end{abstract}


\maketitle


Discreteness effects are known to stabilize surface modes
in nonlinear lattices~\cite{Kivshar:1998-125:PD, Makris:2005-2466:OL}, and stable light
self-trapping at the edge of a nonlinear self-focusing lattice
accompanied by the formation of a discrete surface soliton has
recently been demonstrated experimentally~\cite{Suntsov:2006-63901:PRL}. Such
{\em unstaggered discrete surface modes} can be treated as discrete
solitons trapped at the edge of a waveguide array when the beam
power exceeds a certain critical value associated with a strong
repulsive surface energy~\cite{Molina:2006-discrete:OL}.

On the other hand, staggered linear surface modes are known as Tamm
states~\cite{Tamm:1932-849:ZPhys}, and they were first found in
solid state physics as localized electronic states at the edge of a
truncated periodic potential; an optical analog of linear Tamm
states has been demonstrated for the case of an interface separating
periodic and homogeneous dielectric media~\cite{Yeh:1978-104:APL}.
>From that perspective, staggered surface gap solitons in defocusing
semi-infinite periodic media, recently introduced
theoretically~\cite{Kartashov:2006-73901:PRL} and observed
experimentally~\cite{Rosberg:physics/0603202:ARXIV}, provide a full
analogy to localized electronic surface Tamm states being an optical
realization of {\em nonlinear Tamm states}.

The aim of this Letter is twofold. First, we clarify important links
between different types of linear and nonlinear surface modes and
demonstrate that the analysis of linear surface states in waveguide
arrays with defects provides important information about possible
existence of nonlinear surface modes. Second, we study discrete
nonlinear surface states in semi-infinite binary waveguide arrays,
previously introduced theoretically~\cite{Sukhorukov:2002-2112:OL,
Sukhorukov:2003-2345:OL} and then studied
experimentally~\cite{Morandotti:2004-2890:OL}, and find numerically
the families of both discrete surface solitons and nonlinear Tamm
(gap) states and analyze their stability.

\begin{figure}[htbp]
\centerline{\scalebox{.5}{\includegraphics{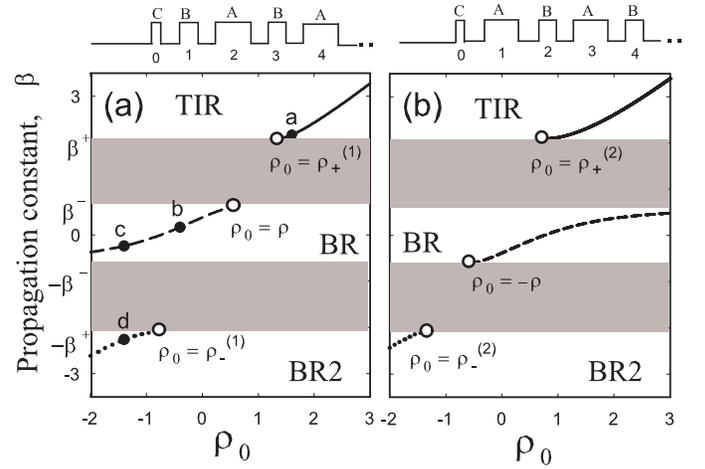}}}
\caption{(a,b)~Spectra of the
linear surface modes in a binary array of alternating thick (A) and
thin (B) optical waveguides with a surface defect (C) for two
configurations of a semi-infinite structure shown schematically on
top. Shaded areas mark two bands of an infinite binary array ($\rho
= 0.6$). Marked points correspond to the examples shown in
Figs.2(a-d).} \label{fig:1}
\end{figure}


We consider propagation  and localization of light in a
semi-infinite periodic binary array of alternating wide and narrow
weakly coupled optical waveguides, as shown schematically in
Fig.~\ref{fig:1}(top). In infinite binary waveguide
arrays, the properties of spatial discrete solitons can be
effectively managed by controlling the geometry of the
array~\cite{Sukhorukov:2002-2112:OL, Sukhorukov:2003-2345:OL,
Morandotti:2004-2890:OL}. Following the earlier
analysis~\cite{Sukhorukov:2002-2112:OL, Sukhorukov:2003-2345:OL}, we
describe the binary array within the tight-binding approximation,
where the total field is decomposed into a superposition of weakly
overlapping modes of the individual waveguides of two kinds (A and
B; wide and narrow). The corresponding equation for the mode
amplitudes takes the form~\cite{Sukhorukov:2002-2112:OL}
\begin{equation}
   i \frac{d E_n}{dz}
   + \rho_n E_n
   + (E_{n-1} + E_{n+1})
   + \gamma_n |E_n|^2 E_n
   = 0,\label{DNLS}
\end{equation}
where $n = 0, 1, \ldots$, and $E_{-1} \equiv 0$ due to the structure
termination. Here $\rho_n$ characterizes the linear propagation
constant of the mode guided by the $n-$th waveguide, $\gamma_n$ are
the effective nonlinear coefficients (we consider Kerr-type medium
response). For the structure shown in Fig.~\ref{fig:1}(a,
top), we have  $\rho_{2 n + 1} = - \rho$, $\rho_{2 n}=\rho$, and for
Fig.~\ref{fig:1}(b, top), $\rho_{2 n + 1} = \rho$,
$\rho_{2 n} = -\rho$, where $\rho>0$ defines the difference between
the propagation constants of the modes guided by wide and narrow
waveguides.

According to Eq.~(\ref{DNLS}), the linear Bloch-wave dispersion is
defined as $K_b = \cos^{-1}( -\eta / 2)$, where $\eta = 2 + \rho^2 -
\beta^2$. The transmission bands correspond to real $K_b$, and they
appear when $\beta_- \le |\beta| \le \beta_+$, where $\beta_- =
|\rho|$ and $\beta_+ = (\rho^2 + 4)^{1/2}$. The band-gap structure is
presented in Figs.~\ref{fig:1}(a,b). The upper gap at
$\beta>\beta_+$ is due to the effect of  total internal reflection
(TIR gap), whereas additional BR and BR2 gaps appear for $|\beta|<\beta_-$
and $\beta<-\beta_+$, respectively, due to Bragg reflection.

In a general case, we assume that the edge waveguide (C) differs
from both broad (A) and narrow (B) waveguides, and it is described
by a different normalized propagation constant, $\rho_{0}$. Then, we
study linear surface modes for two possible configurations of the
binary array with the defect at the edge [see top insets in
Figs.~\ref{fig:1}(a,b)]. We look for the localized
solutions of Eq.~(\ref{DNLS}) at $\gamma_n \equiv 0$ in the form $E_n
= u_n \exp(i \beta z)$, where $\beta$ is the propagation constant
and $u_n$ is the mode profile. Such eigenmodes of the semi-infinite
periodic structure with a surface defect can be found as truncated
Bloch waves of the corresponding infinite structure, where the
equation for the edge waveguide defines an effective boundary
condition, $\rho_0 = -\rho_1 + [1+\exp(- i K_b)] / (\beta -
\rho_1)$. The mode becomes exponentially localized inside the gaps
where the Bloch wave number is complex, and ${\rm Im} K_b(\beta) >
0$. Indeed, we find that spatially localized modes can appear in the
semi-infinite total internal reflection (TIR) gap [solid curve in
Figs.~\ref{fig:1}(a,b)] or inside the Bragg reflection
gaps [dashed and dotted lines in Figs.~\ref{fig:1}(a,b)
correspond to BR and BR2 gaps, respectively].

\begin{figure}[t]
\centerline{\scalebox{.3}{\includegraphics{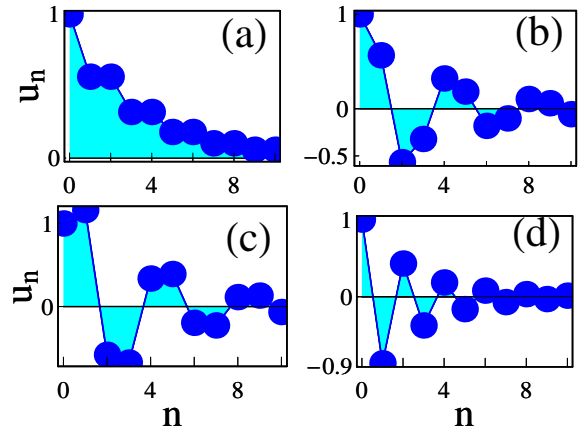}}}
\caption{(Color online) Examples
of the linear surface modes in the binary waveguide array shown in
Fig.~\ref{fig:1}(a). The mode profiles in (a), (b), (c)
and (d) are associated with different gaps of the linear spectrum
and correspond to the points ``a", ``b", ``c" and ``d"  marked in
Fig.~\ref{fig:1}(a).} \label{fig:2}
\end{figure}

We find that the surface localized modes may appear in all three
gaps of the wave spectra, as shown in
Figs.~\ref{fig:1}(a,b), where the marked values of
$\rho_{o}$ are: $\rho_{\pm}^{(1)} = (\rho/2)\pm
\sqrt{(\rho/2)^2+1}$, for the structure in
Fig.~\ref{fig:1}(a) and $\rho_{\pm}^{(2)} = -(\rho/2)\pm
\sqrt{(\rho/2)^2 +1})$, for the structure in
Fig.~\ref{fig:1}(b). Examples of the linear surface modes
from different gaps corresponding to the binary array shown in
Fig.~\ref{fig:1}(a) are presented in
Figs.~\ref{fig:2}(a-d).

One of the major observations is that, for the case when there is no
surface defect and the binary array is just cut, i.e. $\rho_{0} =
\rho$ [see Fig.~\ref{fig:1}(a)] or $\rho_{0} = -\rho$
[see Fig.~\ref{fig:1}(b)], similar to the original
formulation of the problem for the Tamm
states~\cite{Tamm:1932-849:ZPhys}, {\em no linear localized surface
modes} exist in the system, similar to the case of an array composed
of identical waveguides discussed
earlier~\cite{Makris:2005-2466:OL}. The existence of linear surface
modes becomes possible solely due to the presence of a defect
located at the edge of the waveguide array, and such localized modes
are associated with a specific shift of the beam propagation
constant when $\rho_{0} \neq \rho$ [Fig.~\ref{fig:1}(a)]
and $\rho_{0} \neq -\rho$ [Fig.~\ref{fig:1}(b)]. The
structure of a particular surface mode supported by a defect
reflects its location with respect to the photonic bandgap of the
binary array: i.e. it is unstaggered in the  TIR gap
[Fig.~\ref{fig:2}(a)], staggered in the BR gap
[Figs.~\ref{fig:2}(b,c)], and doubly-staggered in the BR2
gap [Fig.~\ref{fig:2}(d)].

As is well established in many problems of nonlinear optics, a
nonlinear dependence of the optical refractive index on the
intensity of the incoming light provides an important physical
mechanism of the intensity-induced shift of the propagation constant
which can act as an effective surface defect. Thus, when we consider
the case of a nonlinear binary waveguide array, the mode propagation
constant becomes shifted by nonlinearity, and for the focusing
nonlinear response this shift will be always positive. We analyze
below two cases where either a wide or a narrow waveguide is placed
at the edge of the array, i.e. when the defect waveguide is created
solely by the nonlinearity, similar to the case of nonlinear
truncated lattices studied earlier~\cite{Makris:2005-2466:OL}. Just
as in the linear case, we look for spatially localized solutions of
Eq.~\ref{DNLS}, now with the nonlinear term (we neglect the
differences in the effective nonlinear coefficients at the A and B
sites and assume $\gamma_n \equiv 1$), in the form $E_n = u_n \exp(i
\beta z)$, where $\beta$ is the propagation constant, and the
function $u_n$ describes the soliton profile.


\begin{figure}[htbp]
\centerline{\scalebox{.31}{\includegraphics{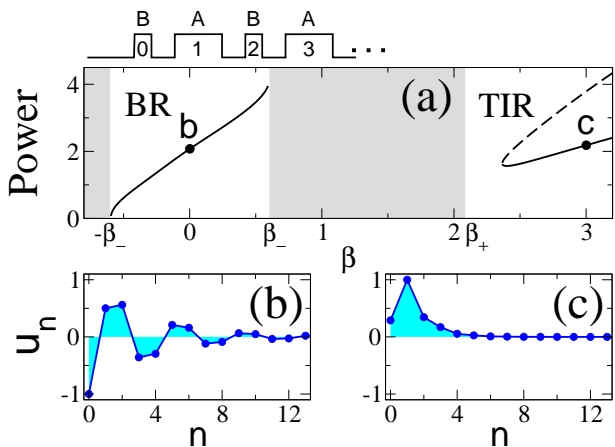}}}
\caption{(Color online) (a)
Families of nonlinear Tamm (gap) states and discrete surface
solitons in the binary array truncated at the narrow waveguide.
Solid and dashed curves correspond to the stable and unstable
branches, respectively. Schematic of the array is shown on the top.
(b,c) Characteristic examples of the soliton profiles corresponding
to the points "b" and "c" marked in (a). Detuning parameter of the
array is $\rho =0.6$. } \label{fig:3}
\end{figure}


\begin{figure}[htbp]
\centerline{\scalebox{.31}{\includegraphics{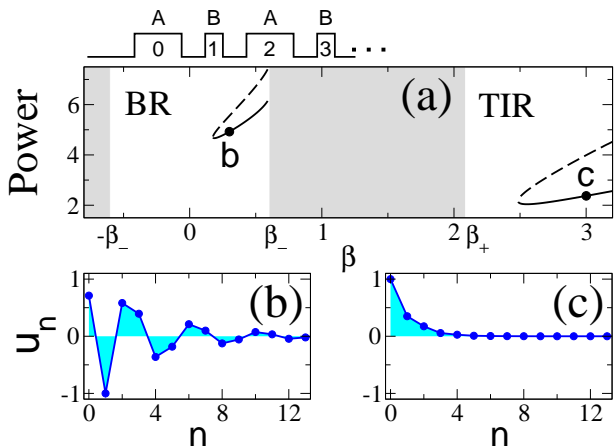}}}
\caption{(Color online) Same as in
Fig.~\ref{fig:3}, but for the binary array truncated at
the wide waveguide. } \label{fig:4}
\end{figure}

In Figs.~\ref{fig:3} and~\ref{fig:4}, we show
the families of nonlinear surface solitons for two kinds of
termination. In the case  when the edge waveguide is narrow
[Fig.~\ref{fig:3}], the nonlinear Tamm states (i.e.
nonlinear surface modes in the BR gap) appear at low power, in
agreement with the linear analysis predicting the existence of
surface state for a small change of the propagation constant [dashed
curve in Fig.~\ref{fig:1}(b)]. On the other hand, the
existence of the surface mode in the TIR gap requires the mode power
to exceed some threshold value, similar to the critical value of the
defect strength found in the linear theory [solid curve in
Fig.~\ref{fig:1}(b)]. When the binary lattice is
truncated at the wide waveguide [Fig.~\ref{fig:4}], a
finite power is required to support the surface solitons in either
TIR or BR gaps, reflecting the properties of linear TIR modes which
can be supported only when a positive shift of the propagation
constant exceeds a certain threshold value [solid and dashed curves
in Fig.~\ref{fig:1}(a)].

Next, we employ the beam propagation method to study the soliton
stability. In Figs.~\ref{fig:3}(a)
and~\ref{fig:4}(a), dashed branches of the curves indicate
unstable surface modes. Whereas oscillatory
instabilities~\cite{Pelinovsky:2004-36618:PRE} may arise for the
solid branches, we have verified that the solitons corresponding to
the marked points on the solid curves demonstrate stable propagation
for more than 100 coupling lengths even in the presence of initial
perturbations.

Finally, we mention that the recent studies of discrete gap solitons
in binary waveguide arrays provided an experimental
evidence~\cite{Morandotti:2004-2890:OL} of a controlled soliton
generation  and power-dependent soliton steering in  agreement with
earlier theoretical predictions~\cite{Sukhorukov:2002-2112:OL,
Sukhorukov:2003-2345:OL}. The surface modes in the form of discrete
surface solitons and nonlinear Tamm (gap) states described in this
paper extend the class of such discrete nonlinear modes, and they
can be treated as the discrete solitons trapped at the edge of a
waveguide array when the beam power exceeds a certain critical value
associated with a strong repulsive surface
energy~\cite{Molina:2006-discrete:OL}. We believe that our results
open a road for the experimental studies of intriguing properties of
discrete surface gap solitons in engineered periodic structures.

In conclusion, we have analyzed linear surface states at the edge of
a semi-infinite binary array of optical waveguides, and revealed a
link between the linear surface modes supported by the defects and
nonlinear localized states in the defect-free semi-infinite
structure. We have found families of both discrete surface solitons
and nonlinear Tamm states and analyzed their stability.

This work has been supported by Fondecyt grants 1050193 and
7050173 in Chile, and by the Australian Research Council in
Australia.


\begin{thebibliography}{10}

\bibitem{Kivshar:1998-125:PD}
Yu.~S. Kivshar, F. Zhang, and S. Takeno, Physica D {\bf 119}, 125
(1998).

\bibitem{Makris:2005-2466:OL}
K.~G. Makris, S. Suntsov, D.~N. Christodoulides, G.~I. Stegeman, and
A. Hache,
  Opt. Lett. {\bf 30}, 2466 (2005).

\bibitem{Suntsov:2006-63901:PRL}
S. Suntsov, K.~G. Makris, D.~N. Christodoulides, G.~I. Stegeman, A.
Hache, R.
  Morandotti, H. Yang, G. Salamo, and M. Sorel, Phys. Rev. Lett. {\bf 96},
  063901 (2006).

\bibitem{Molina:2006-discrete:OL}
M. Molina, R. Vicencio, and Yu.~S. Kivshar, Opt. Lett. {\bf 31}
(2006), in
  press.

\bibitem{Tamm:1932-849:ZPhys}
I.~E. Tamm, Z. Phys. {\bf 76}, 849 (1932).

\bibitem{Yeh:1978-104:APL}
P. Yeh, A. Yariv, and A.~Y. Cho, Appl. Phys. Lett. {\bf 32}, 104
(1978).

\bibitem{Kartashov:2006-73901:PRL}
Y.~V. Kartashov, V.~A. Vysloukh, and L. Torner, Phys. Rev. Lett.
{\bf 96},
  073901 (2006).

\bibitem{Rosberg:physics/0603202:ARXIV}
C.~R. Rosberg, D.~N. Neshev, W. Krolikowski, Yu.~S. Kivshar, A.
Mitchell, R.~A.
  Vicencio, and M.~I. Molina, arXiv {\bf {\mdseries physics/0603202}} (2006).

\bibitem{Sukhorukov:2002-2112:OL}
A.~A. Sukhorukov and Yu.~S. Kivshar, Opt. Lett. {\bf 27}, 2112
(2002).

\bibitem{Sukhorukov:2003-2345:OL}
A.~A. Sukhorukov and Yu.~S. Kivshar, Opt. Lett. {\bf 28}, 2345
(2003).

\bibitem{Morandotti:2004-2890:OL}
R. Morandotti, D. Mandelik, Y. Silberberg, J.~S. Aitchison, M.
Sorel, D.~N.
  Christodoulides, A.~A. Sukhorukov, and Yu.~S. Kivshar, Opt. Lett. {\bf 29},
  2890 (2004).

\bibitem{Pelinovsky:2004-36618:PRE}
D.~E. Pelinovsky, A.~A. Sukhorukov, and Yu.~S. Kivshar, Phys. Rev. E
{\bf 70},
  036618 (2004).

\end{thebibliography}
\end{document}